# Wearable proximity sensors for monitoring a mass casualty incident exercise: a feasibility study

Laura Ozella[1*], Laetitia Gauvin[1], Luca Carenzo[2,4], Marco Quaggiotto[1,3], Pier Luigi Ingrassia[4], Michele Tizzoni[1], André Panisson[1], Davide Colombo[2], Anna Sapienza[5,1], Kyriaki Kalimeri[1], Francesco Della Corte[2], Ciro Cattuto[1]

1. Data Science Laboratory, ISI Foundation, Turin, Italy.

2. CRIMEDIM Research Centre in Emergency and Disaster Medicine, Università del Piemonte Orientale, Novara, Italy

3. Department of Design, Politecnico di Milano, Milano, Italy

4. Centro Interdipartimentale di Didattica Innovativa e di Simulazione in Medicina e Professioni Sanitarie" SIM-NOVA, Università del Piemonte Orientale, Novara, Italy

5. USC Information Sciences Institute, Marina del Rey, CA 90292, USA

*Corresponding author. E-mail: laura.ozella@isi.it

## Abstract

**Background:** Over the past several decades, naturally occurring and man-made mass casualty incidents (MCI) have increased in frequency and number, worldwide. To test the impact of such event on medical resources, simulations can provide a safe, controlled setting while replicating the chaotic environment typical of an actual disaster. A standardised method to collect and analyse data from mass casualty exercises is needed, in order to assess preparedness and performance of the healthcare staff involved.

**Objective:** We report on the use of wearable proximity sensors to measure proximity events during a MCI simulation. We investigated the interactions between medical staff and patients, to evaluate the time dedicated by the medical staff with respect to the severity of the injury of the victims depending on the roles. Moreover, we estimated the presence of the patients in the different spaces of the field hospital, in order to study the patients' flow.

**Methods:** Data were obtained and collected through the deployment of wearable proximity sensors during a mass casualty incident functional exercise. The scenario included two areas: the accident site and the Advanced Medical Post (AMP), and the exercise lasted 3 hours. A total of 238 participants were involved in the exercise and classified in categories according to their role: 14 medical doctors, 16 nurses, 134 victims, 47 Emergency Medical Services staff members, and 27 healthcare assistants and other hospital support staff. Each victim was assigned a score related to the severity of his injury. Each participant wore a proximity sensor and, in addition, 30 fixed devices were placed in the field hospital.

**Results:** The contact networks show a heterogeneous distribution of the cumulative time spent in proximity by participants. We obtained contact matrices based on cumulative time spent in proximity between victims and the rescuers. Our results showed that the time spent in proximity by the healthcare teams with the victims is related to the severity of the patient's injury. The analysis of patients' flow showed that the presence of patients in the rooms of the hospital is consistent with triage code and diagnosis, and no obvious bottlenecks were found.

**Conclusions:** Our study shows the feasibility of the use of wearable sensors for tracking close contacts among individuals during a mass casualty incident simulation. It represents, to our knowledge, the first example of unsupervised data collection of face-to-face contacts during a MCI exercise. The unsupervised measurement of contact patterns with proximity sensors provides an unique opportunity to monitor the interactions between participants without the involvement of direct observers, which could compromise the realism of the exercise. Moreover, the use of the sensors as fixed devices allowed to analyse the flow of the patients in the field hospital, in order to assess if they were optimally headed by the healthcare personnel.

# Introduction

## Background

A mass casualty incident (MCI) is defined as a situation in which, at a certain time, the available care resources are unable to meet the demand for medical care of the incident [1]. Each year MCIs occur worldwide, and are caused by conventional causes such as weapons, explosions, vehicular and airplane accidents and deliberate or spontaneous chemical mass intoxications. These incidents require emergency healthcare teams to treat large numbers of injured victims [2], and this might compromise the normal functioning of hospitals. Simulation applied to healthcare is rapidly gaining acceptance in medical and academic communities and it can be a valuable tool for better training of management of MCIs. It provides a safe, controlled environment in which is possible to test plans and procedures and improve them, as well as to evaluate policies and guidelines [3]. Traditionally, actors are used in disaster exercises, and they are coached to mimic and to exhibit realistic manifestations of several medical and traumatic pathologic states that may present in a real MCI [2-4]. Although the use of simulation in medical education has increased over the last two decades, collecting and analysing data of a mass casualty functional exercise still happens in an unsystematic manner, without a standardised method. Common methods to assess performances during MCI simulations are direct observations of functional exercise performance, and video analysis of participants' behaviours [5]. However, these methods present some limitations: the focus of the observers' attention can subjectively vary, and need of several observers both for direct observations and for videos could affect the realism of the event and decrease the level of the emotional engagement of the participants [3]. An objective and reproducible method that identifies the strengths and weaknesses of simulations is required in order to lead the improvement in the response system [6]. In mass casualty simulations, wireless medical sensor networks and Radio-Frequency IDentification (RFID) technology have been used to track information about the status of the casualties, thus providing timely situational awareness during exercises [3, 8, 9], and the use of RFID was compared with manual data collection, demonstrating the reliability and applicability of the system [3]. Wearable proximity devices could provide not only patient information and tracking capability to locate people and equipment, but also information on interactions among individuals. Wearable sensors have been successfully used to measure face-to-face proximity relations in various hospital settings that include pediatric ward [10], and acute care geriatric unit [11, 12].

## Goals of this study

The use of proximity sensors to the field of MCI simulation could provide a continuous and fully distributed collection system of high-resolution data on the interactions among patients and medical staff, in order to investigate the dynamics of interactions with respect to the different roles and severity of the patient's injuries. Here, we illustrate the feasibility of contact measures through wearable proximity sensors in a live MCI simulation, aimed at providing data-driven knowledge to perform debriefing and identify improvement. The main objectives of our work are (1) to investigate the interactions between medical staff and patients, with respect to their roles and severity of the victim's condition; (2) to estimate the presence of victims in different spaces of the field hospital, in order to study the patients' flow.

# Methods

## Study setting

A building collapse following a flood was simulated during a MCI functional exercise organised in Novara, Italy on 19th May 2016 from 07:00 PM to 10:30 PM. The MCI exercise was organised in the framework of the residential course of the European Master in Disaster Medicine (EMDM). The EMDM is an international 12-month-long blended learning master degree program for healthcare providers involved in medical preparedness and response to disasters [13]. The exercise included both a pre-hospital and an in-hospital disaster response phase. The scenario comprised two locations: the building collapse site (pre-hospital response) and the field hospital (in-hospital response). The hospital was located approximately two kilometers from the incident site. Overall, exercise participants were distinguished into the following classes for the purpose of the present study and based on their role in the simulation: Medical Doctors ('MD'), Nurses ('Nurse') Emergency Medical Services personnel ('EMS'), and Healthcare Assistants ('HCA') and simulated victims ('Victim'). Victims were portrayed by medical students. They attended an introductory course on disaster medicine (8-hour live lectures) and specific training on how to simulate clinical conditions provided in an individual victim storyboard, when to change the dynamic casualty cards reporting their vitals (DCCs) according to the treatment applied, and how to properly collect data (2-hour live lectures). Details about casualty evolution method, general structure of the simulation and DCCs were described in a series of previous papers [3, 6, 7, 14]. EMDM students acted as doctors and nurses and were distributed as follows: 8 physicians and 5 nurses staffed the ambulances provided by local EMS Agencies as pre-hospital response whilst 6 physicians and 11 nurses were in the field hospital which had been previously kindly deployed by the Italian Army. EMS personnel and HCA were played by local ambulance volunteers (basic emergency medical technician level) and first aid trained soldiers respectively. None of the participants had been previously informed about the scenario.

## Expected Triage and Injury Severity Score

According to their predetermined storyboard, each victim had an expected initial triage category according to the START protocol [15]: 6 victims were *Black*, 15 *Red*, 27 *Yellow*, and 86 *Green*. Responders had to assign a triage score to each victim during the exercise, both at the accident site ('on-scene triage') and at the hospital ('hospital triage').

Victims (*Black* group excluded) were also classified based on their injuries using the Injury Severity Score (ISS) [16]. The ISS is an established medical score to assess trauma severity with a range from 1 to 75, grouped by five categories: *Minor* (1-3), *Moderate* (4-8), *Serious* (9-15), *Severe* (16-24), and *Critical* (25-75). In addition, the category *Non-traumatic*, which indicates victims without physical trauma (like anxiety crises), was added. In total 21 victims were assigned to *Non-traumatic* group, 36 to *Minor* group, 47 to *Moderate* group, 9 to *Serious* group, 4 to *Severe* group, and 11 to *Critical* group.

## Data collection

Data collection was performed as previously described. Each participant wore a wearable proximity sensor: the sensor was inserted into a transparent envelope, and fixed with adhesive tape at the centre of the chest (on the sternum area) in order to detect person-to-person interactions. At the beginning of the simulation, victims were both at the accident site and at the hospital as regular in-hospital patient. During the exercise, victims could take one of three possible pathways: (1) transferred from the accident sites to the hospital, by ambulances and minibus (2) transferred to another virtual hospital (3) discharged from the simulation. The exact times of the transfers as well as the ending time of the simulation for each victim were marked by external observers. In addition, proximity sensors were placed on the ceiling of the rooms (tents) of the Hospital area (Figure 1) (category 'Location') as fixed tags. Table 1 reports a summary of the total number of sensors for each category, on Pre-hospital and Hospital area.

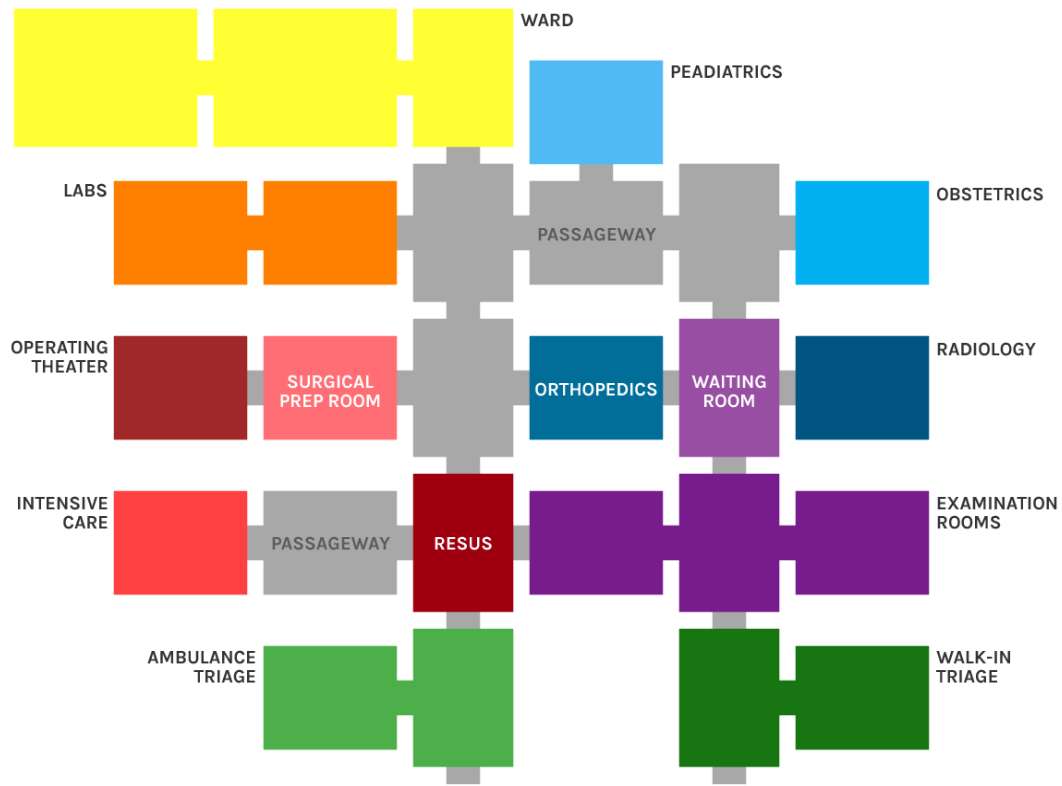

Figure 1. Map of the field hospital.

Table 1. Number of proximity sensors, on Pre-hospital and Hospital area by category.

|  | *Pre-hospital* | *Hospital* | Total |
|---|---|---|---|
| **MD** | 8 | 6 | 14 |
| **Nurse** | 5 | 11 | 16 |
| **Victim** | 112 | 22 | 134 |
| **EMS** | 47 |  | 47 |
| **HCA** |  | 27 | 27 |
| **Location** |  | 30 | 30 |
| **Total** | 172 | 96 | **268** |

The data were processed using a proximity-sensing platform developed by the SocioPatterns collaboration consortium [17]. This system is based on wearable proximity sensors ('tags') that exchange ultra-low-power radio packets in a peer-to-peer fashion [10, 18, 19, 20]. Sensors in close proximity exchange with one another a maximum of about 1 power packet per second, and the exchange of low-power radio-packets is used as a proxy for the spatial proximity between tags [10, 18]. In particular, close proximity is measured by the attenuation, defined as the difference between the received and transmitted power. Each device has a unique identification number that was used to link the information on the individual carrying the device with his/her profile, or in case of fixed tags, with the location where the sensors are placed.

## Contacts among participants

We analysed the contacts among participants belonging to categories 'Victim', 'MD', 'Nurse', and 'EMS' for the Pre-hospital area, and the contacts between participants belonging to categories 'Victim', 'MD', 'Nurse', and 'HCA' for the Hospital area. We considered the contacts between individuals across categories, both in Pre-hospital and Hospital area. We defined that a 'contact' occurs between two individuals, during a time slice duration of 20 seconds, if and only if the proximity devices worn by the participants exchanged at least one radio packet during that interval. After a contact is established, it is considered ongoing as long as the devices continue to exchange at least one such packet for every subsequent 20 s interval [18]. The system was set in order to detect proximity events between devices situated within 1-1.5 m of one another. This setting ensures that when individuals wear the devices on their chest, exchange of radio packets between devices is only possible when they are facing each other, as the human body acts as an RF shield at the carrier frequency used for communication. This system allows us to monitor the number of contacts and their duration. Data were extracted and cleaned separately for each sensor, and those collected before 07:00 PM and after 10:30 PM were discarded to keep track of only meaningful proximity events. Moreover, for each victim, we discarded the data collected after the exact time of simulation end (death, discharge or end-of-simulation time). We analysed the data separately for the Pre-hospital and Hospital area. As regards the victims transferred from the accident site to Hospital during the simulation, we considered the data collected before the exact time of transfer belonging to Pre-hospital data, and data collected after this time belonging to Hospital data.

We generated aggregated networks of contacts between participants on the full exercise duration, both in Pre-hospital, and in Hospital area, in order to study the close-range interactions during the exercise as well as to confront the results with those obtained in different real-world settings. We call $k_i$ the degree of a node $i$, *i.e.*, the number of distinct individuals with whom individual $i$ has been in contact, and $w_{ij}$, the weight of an edge between nodes $i$ and $j$, *i.e.*, the cumulative duration of the contact events recorded between two individuals $i$ and $j$.

Then, we generated contact matrices based on the median cumulative time spent in proximity between victims with different triage and ISS, and the caregivers (medical doctors, nurses, emergency medical services, and healthcare assistants). Time spent in proximity with victims for each caregivers' category was compared using the Kruskal–Wallis test. We respectively considered the 'on-scene triage' scores and the 'hospital triage' scores to build the Pre-hospital matrix and the Hospital matrix (*i.e.*, the triage scores assigned by the medical doctors).

## Presence of victims in the field hospital

We estimated the presence of the patients in different rooms of the field hospital by analysing the power packets exchanged between sensors wearing by individuals belonging to the category 'Victim' and the fixed sensors belonging to category 'Location' in Hospital area. In order to assess the location of a patient in a given room at a given time, we set up two thresholds on the count of power packets exchanged between the devices, respectively to evaluate the presence of the participant in the exercise and the presence of the participant in a given room. For each time slots of 5 minutes, we assume that a participant is still present in the exercise site if the total count of the power packets exchanged between all the fixed tags and the participant's tag is greater than 15. We assume that a participant is present in a given room if the total number of power packets exchanged between his tag and the fixed tag of the room considered is higher than 5 for each time slot.

This allows describing the patients' flow through the rooms of the field hospital. The field hospital consisted of 27 rooms organised as follow: 3 general wards, 2 laboratories, 5 passageways (hallways), 3 examination rooms, 1 paediatric ward, 1 waiting room, 1 obstetric ward, 1 operating theatre with two beds, 1 surgical preparation room, 1 orthopaedic ward, 1 radiology waiting room, 1 radiology, 1 intensive care with three beds, 1 emergency department resuscitation (resus) area, 2 ambulance triage rooms (patients brought in by ambulance), and 2 walk-in-triage rooms. In this analysis we grouped the patients' wards, the laboratories, the passageways, the examination rooms, the ambulance triage rooms, and walk-in-triage rooms in the same space.

## Presence patterns of victims in the Hospital

In order to study the link between the presence patterns of victims and the conditions of the victims, we used the t-Stochastic Neighbor Embedding technique (t-SNE) that converts a high-dimensional data set into a matrix of pair-wise similarities and allows to visualize the resulting similarity data [21]. Here, we used as an input dataset a set of vectors where each vector described the spatial features of each victim. More exactly, each victim is initially represented as a vector where elements are time spent by that patient in a given room of the field hospital (normalized on total presence duration).

# Results

## Network analysis and contact among victims and rescuers

A total of 238 individuals participated in the exercise. They were categorized as follows: 14 MD, 16 Nurses, 134 victims, 47 EMS, and 27 HCA. The contacts within the same category were not included in this analysis. Figure 2 shows the degree and the weight distribution in Pre-hospital and Hospital area. The aggregated contact network on Pre-hospital area is formed by 172 nodes and 2035 edges, and the average degree is ⟨ k ⟩ = 23.66 (range 1 - 94), on Hospital area the network is formed by 124 nodes and 1335 edges, and the average degree is ⟨ k ⟩ = 21.53 (range 1 - 58). The weight distribution is heterogeneous in both areas, with heavy-tailed distributions: most contacts are short and there are few long-lasting contacts.

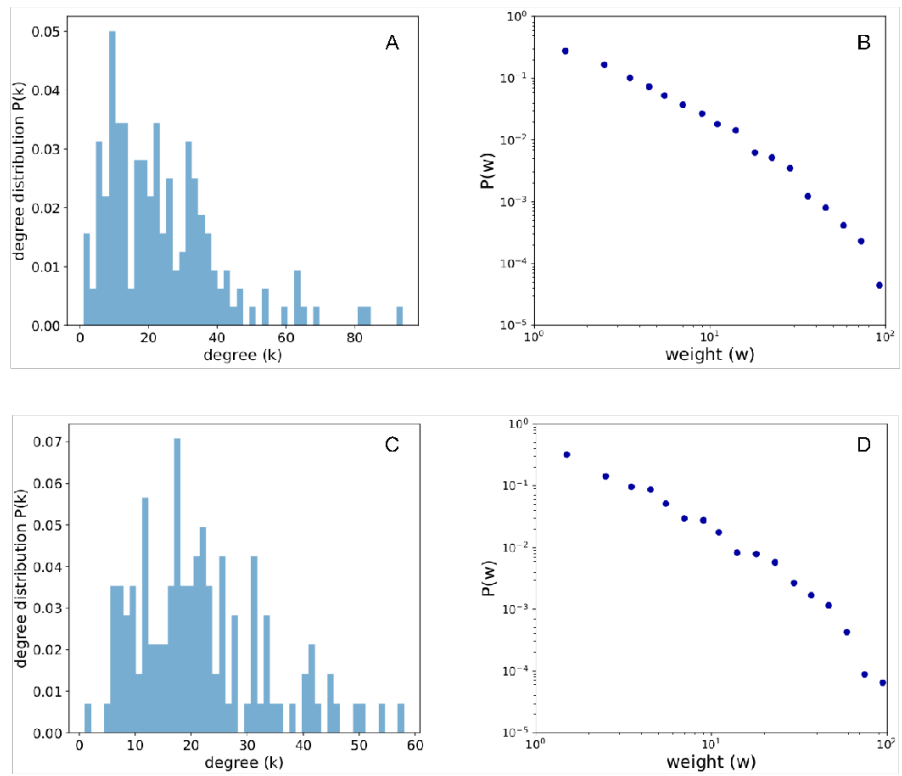

Figure 2: Degree and weight distributions. Degree distribution P(k) of the aggregated contact networks, in Pre-Hospital area (panel A) and in Hospital area (panel C). Distribution of the weights of the aggregated contact

networks in Pre-Hospital area (panel B) and in Hospital area (panel D).

Contact matrices reveal different amount of time spent in proximity depending on the severity of the patient and the role of caregiver (Figures 3 and 4). At the scene of the accident, there was a significant difference between the time spent in proximity between EMS and victims both with respect to the triage ($X^2$ = 19.479, $df$ = 3, $P$ < .001), and to the ISS ($X^2$ =36.106, $df$ = 5, $P$ < .001). The higher time in contact was with Green victims, and with victims classified as Moderate. Regarding the triage, there were no significant differences between the time spent in proximity with the victims for both MD and nurses. On the other hand, regarding the ISS, there were significant difference for MD ($X^2$ = 13.576, $df$ = 5, $P$ = .02), and nurses ($X^2$ = 12.798, $df$ = 5, $P$ = .02). Both categories spent the higher time in contact with victims classified as Moderate and Critical. At the field hospital, there was a significant difference between the time spent in proximity between HCA and victims both with respect to the triage ($X^2$ = 31.271, $df$ = 3, $P$ < .001), and to the ISS ($X^2$ = 46.989, $df$ = 5, $P$ < .001). The higher time in contact was with Green victims, and with victims classified as Minor and Moderate. There were no significant differences between the time spent in proximity between MD and victims (with respect to triage and to ISS) and nurses and victims (with respect to triage). However, there was a significant difference between the time spent in proximity between nurses and victims with respect to ISS ($X^2$ = 14.965, $df$ = 5, $P$ = .01), nurses spent the higher time in contact with victims classified as Moderate and Serious.

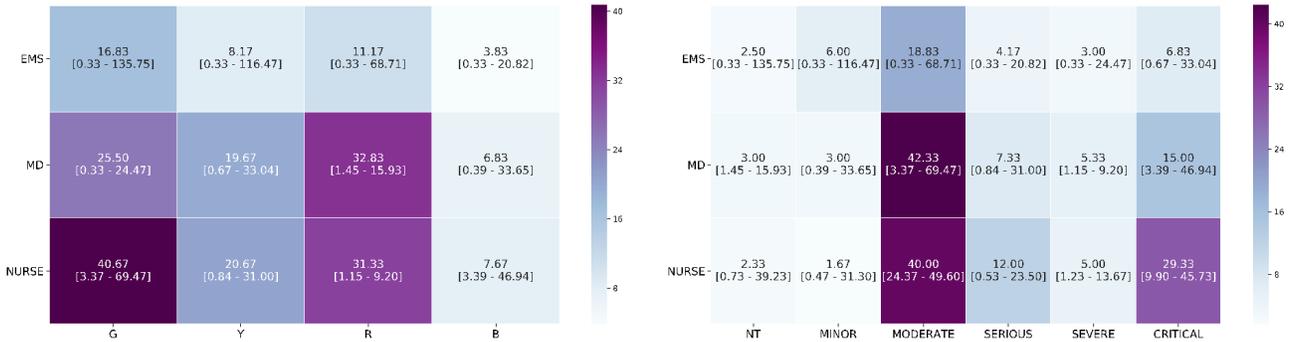

Figure 3. Pre-hospital contact matrices. Median of cumulative time spent (in minutes) between patients with different triage score (G = Green, Y = Yellow, R = Red, B = Black) and rescuers (EMS = Emergency Medical Services, MD = Medical doctors, and Nurses) (left panel); Median of cumulative time spent (in minutes) between patients with different ISS (NT = non-traumatic, Minor, Moderate, Serious, Severe, and Critical) and rescuers (EMS = Emergency Medical Services, MD = Medical doctors, and Nurses) (right panel). 95% confidence intervals are indicated in bracket.

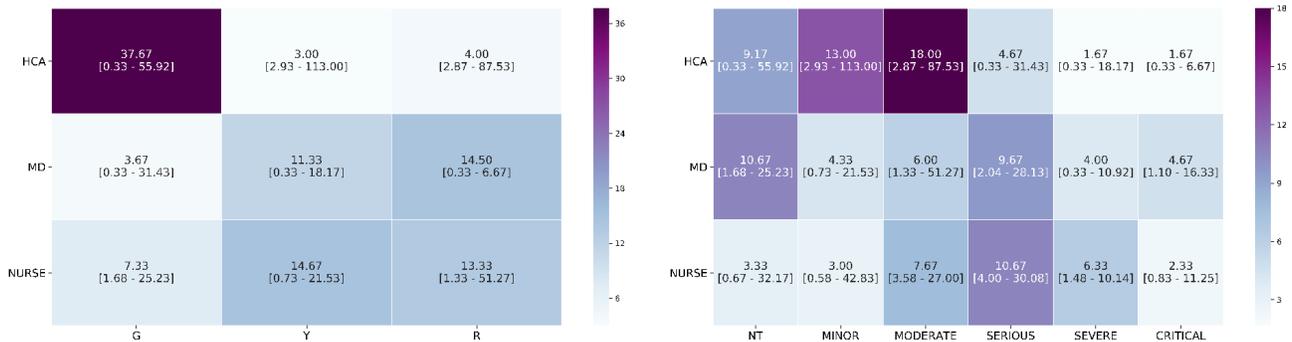

Figure 4. Hospital contact matrices. Median of cumulative time spent (in minutes) between patients with different triage score (G = Green, Y = Yellow, R = Red) and rescuers (HCA = Healthcare Assistants, MD = Medical doctors, and Nurses) (left panel); Median of cumulative time spent (in minutes) between with different ISS (NT = non-traumatic, Minor, Moderate, Serious, Severe, and Critical) and rescuers (HCA = Healthcare Assistants, MD = Medical doctors, and Nurses) (right panel). 95% confidence intervals are indicated in bracket.

Figure 5 shows the cumulative time in contact (normalized on total number of participants belonging to each caregiver category) between caregivers and victims with different triage at Pre-hospital area (panel A) and Hospital area (panel B). The rescuers who spent more cumulative time in contact with victims were nurses at the scene of the accident, and HCA at the field hospital.

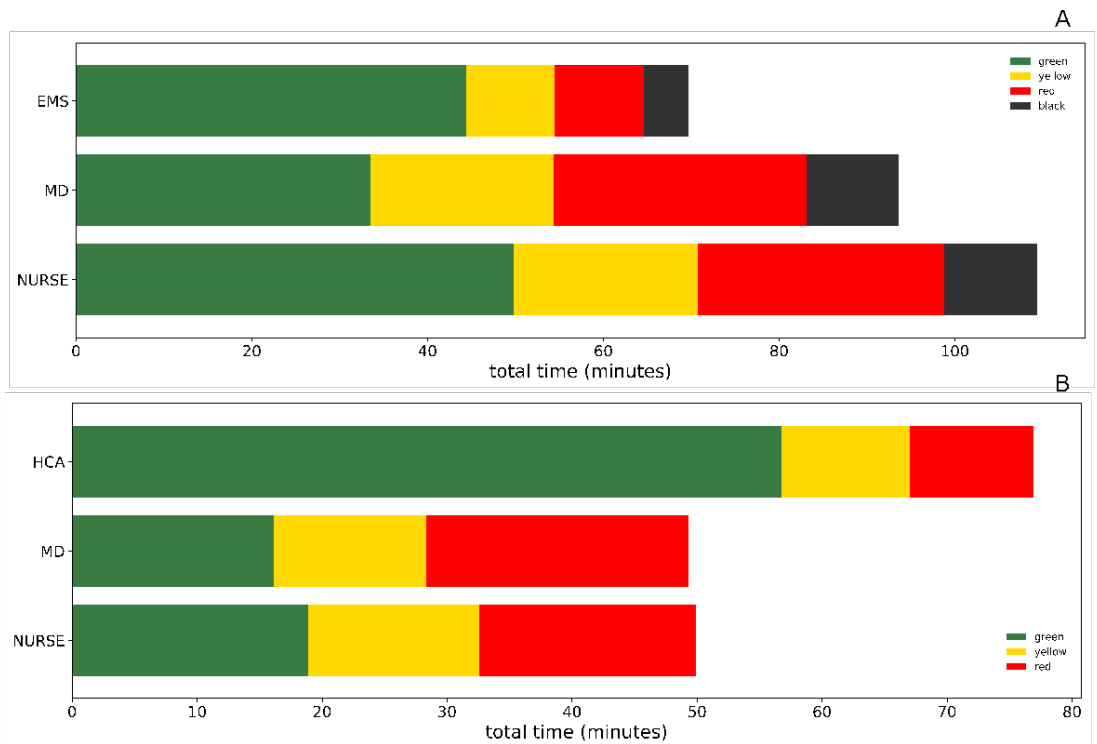

Figure 5. Cumulative time in contact (normalized on the total number of participants belonging to each caregiver category) between caregivers (EMS = Emergency Medical Services, HCA = Healthcare Assistants, MD = Medical doctors, and Nurses) and victims with different triage at Pre-hospital area (panel A) and Hospital area (panel B).

## Casualty flows in the Hospital

We studied the flow of 80 patients: 22 victims were already in the field hospital at the start of simulation as regular in-hospital patients, and 58 victims were transferred from the accident site (the first transfer occurred at 8:15 PM). Figure 6, panel A, shows the victims flow through the field hospital, of 56 patients with Green triage, 14 patients with Yellow, and 10 patients with Red triage. Each bar represents a patient, the colour of the bar's segments refers to the room of the field hospital, and the length of the segments represents time passed by the victim in each room. The presence of patients in the different rooms of the hospital is consistent with triage code and diagnosis. Green victims passed most of their time in Ward, Examination room, Walk-in-triage room and Waiting room. Eleven Yellow patients out of fourteen spent time in Ambulance Triage, six Red victims out of ten spent time in Resus, and three Red victims spent time in the Intensive care. Figure 6, panel B, shows the number of victims in Ward, Ambulance triage, Examination room, Waiting room, and Walk-in-triage over the simulation period. Each line corresponds to the presence of a victim, the colour corresponds to the triage code. By analysing the flow, we aimed at detecting the potential presence of bottlenecks in the field hospital. In order to do this, we focus on the analysis of the presence of the victims in rooms in which they were not receiving any medical treatments. We define bottlenecks as situations where the time of victims spent in the rooms where they did not receive any treatment is increased compared to the average time normally observed (for instance when the number of victims in the hospital is low). Such rooms in the present settings are: Ambulance triage, Examination room, Waiting room, and Walk-in-triage. We compared the numbers of victims present at the same time at the same room in relation to the waiting time in this room, Pearson's correlation test was used. The mean waiting time in Ambulance triage was 40 minutes (SD 9), in Examination room was 40 minutes (SD 9), in Waiting room was 39 minutes (SD 10), and in Walk-in-triage was 31 minutes (SD 23). There were no significant correlations between the number of victims present in the same time at the same room and the waiting time, this result indicates that as the time passed in a room by a patient is not affected by the arrival of many victims in the hospital, in other terms, there were no obvious presence of bottlenecks.

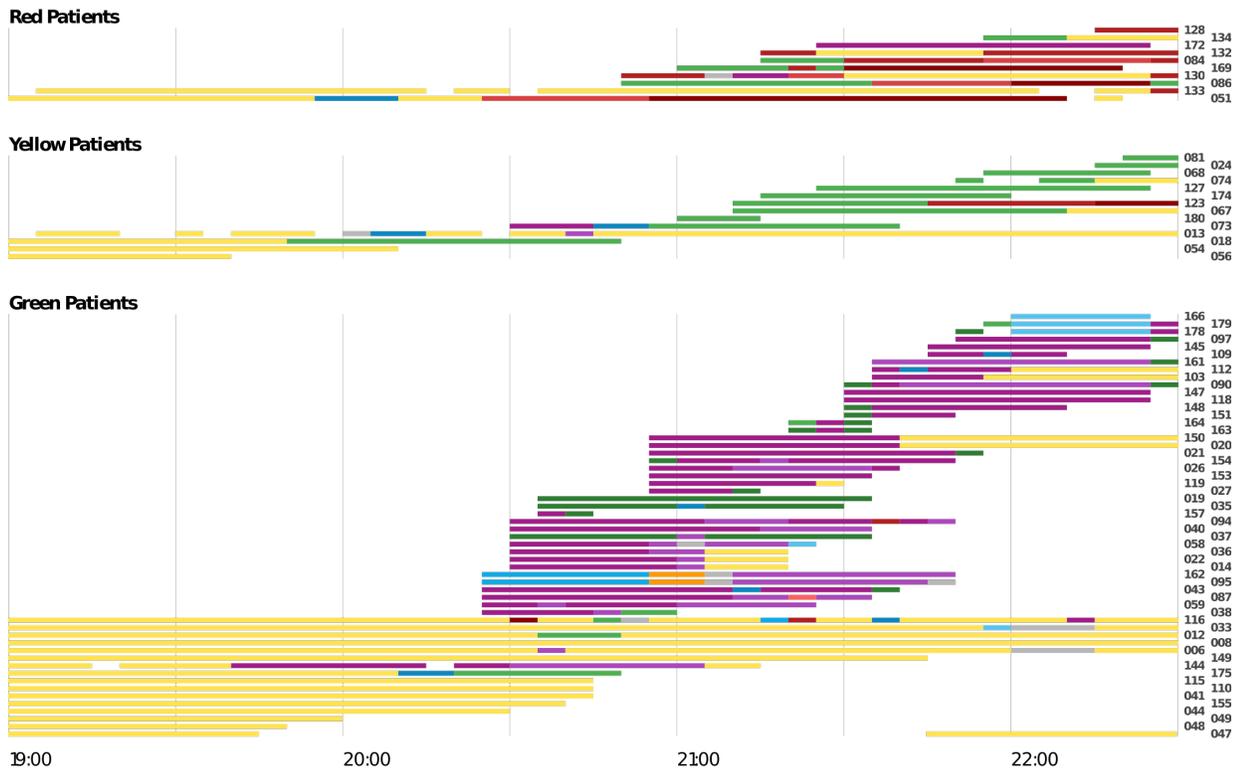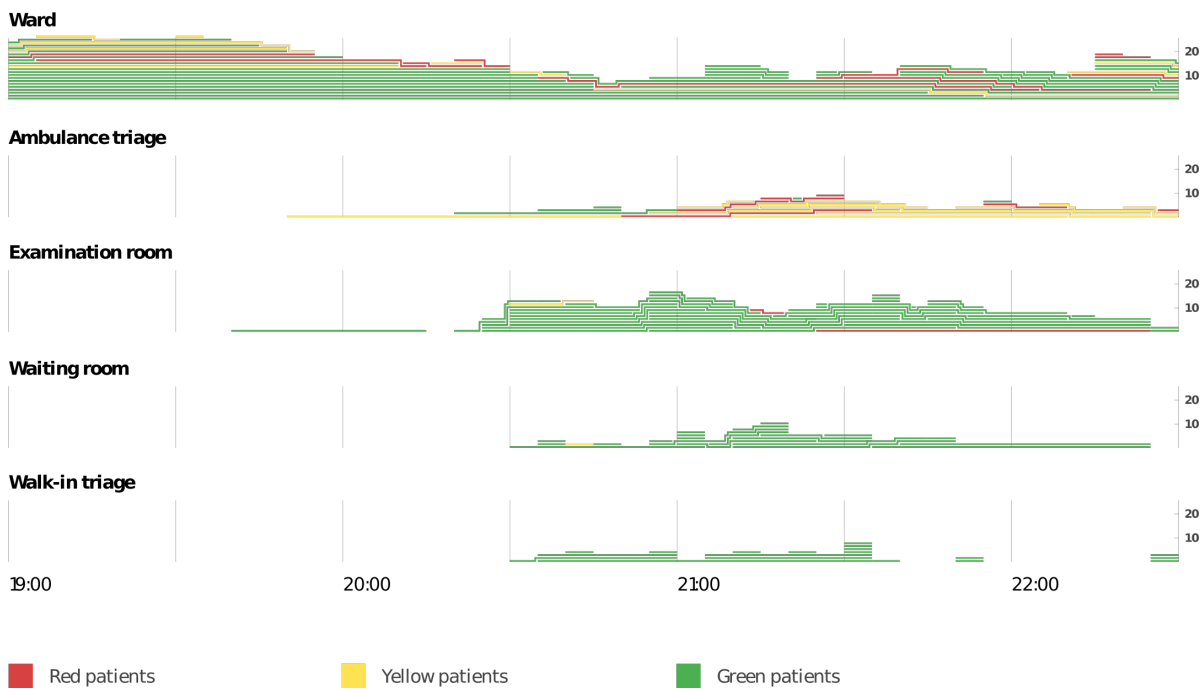

Figure 6. Panel A: Victims flow through the field hospital. Each bar represents a patient (code is indicated on the right of the bar), the colour of the bar's segments refers to the room of the field hospital, and the length of the segments represents time passed by the victim in each room. Numbers on the right part of the figures are identification number for each victim. Panel B: Number of victims in Ward, Ambulance triage, Examination room, Waiting room, and Walk-in-triage over the simulation period. Each line corresponds to the presence of a victim, the colour corresponds to the triage code. Numbers on the right part of the figure indicate the number of people in the room.

## Triage Accuracy

Nine victims expected to be dead on the scene were correctly triaged as black at the scene of the accident. Nine expected red casualties evolved negatively on scene and died before transportation and scene evacuation. Triage accuracy (assigned versus expected) was 85% overall. Triage correctness, over and under-triage stratified per severity is presented in Table 2.

Table 2. Correct and incorrect percentage of assigned triage.

| Triage colour code | Correct (%) | Incorrect | |
| --- | --- | --- | --- |
| | | *Over (%)* | *Under (%)* |
| Red | 100 | 0 | 0 |
| Yellow | 51.86 | 25.92 | 22.22 |
| Green | 95.35 | 4.65 | 0 |
| Black | 84.61 | 0 | 15.39 |
| **Total** | 85.82 | 8.20 | 5.98 |

## Presence patterns of victims in the Hospital

We studied the presence patterns of individual victims in hospital rooms. For each patient we build a feature vector containing the time spent by that patient in each of the 15 hospital rooms normalized by total presence duration. The resulting set of 15-dimensional vectors (one per patient) is visualized using a dimensionality reduction technique known as t-Distributed Stochastic Neighbor Embedding (t-SNE), that maps each 15-dimensional patient vector to a 2-dimensional feature space ($X_1$ and $X_2$ axes) (Figure 7). Clusters of patients with similar presence patterns are visible. On colour coding by assigned triage code, we observe that the bottom-right cluster contains the more serious cases (yellow and red codes), and the up-right cluster contains the less serious cases (green codes), with the exception for the victim coded 172. The ideal triage code of the victim 172 was Yellow, and this victim passed the entire time of the simulation in the Examination room. On colour coding by start location the bottom-left cluster contains the majority of victims that started the simulation on the field hospital.

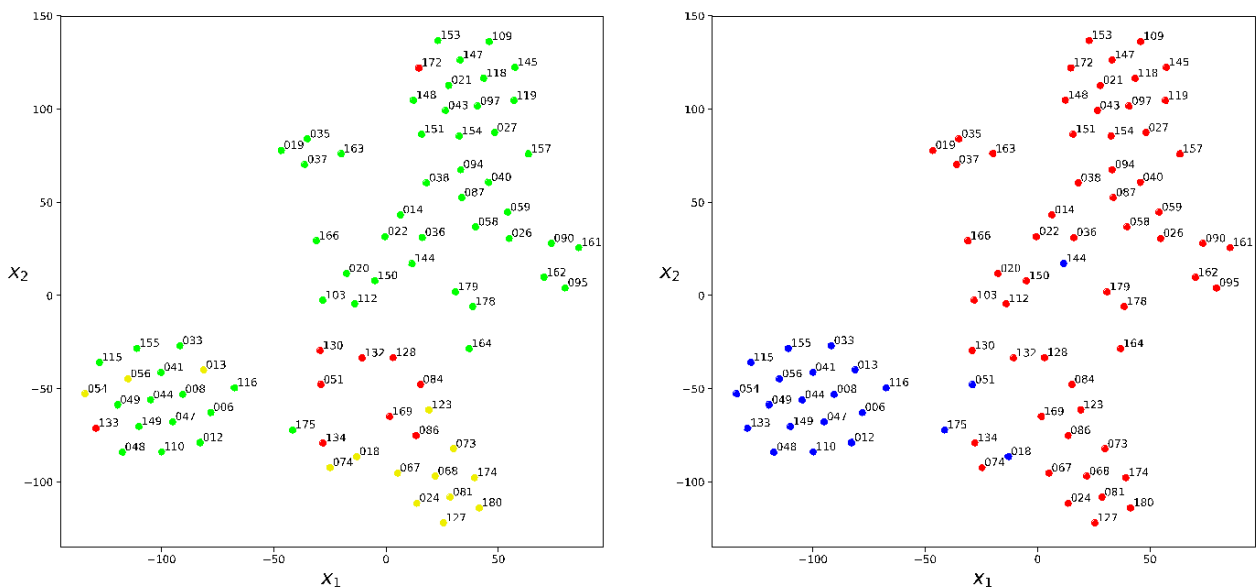

Figure 7. Presence patterns of individual victims in hospital rooms visualized using t-Distributed Stochastic

Neighbor Embedding (t-SNE) to map high-dimensional patient vectors to a 2-dimensional feature space ($X_1$ and $X_2$ axes). Each point corresponds to a patient. Victims with similar presence vectors are mapped to neighboring points in the plane. Victims are colour coding by triage code (left panel), and by the start location of the simulation (red: accident site; blue: field hospital)

Moreover, we studied the presence times of individual victims in three hospital rooms characterized by longer presence times of the victims (Figure 8). We observed that the bottom-left group of patients on the t-SNE plot are characterized by long presence times in the Ward and they correspond to the victims that started the simulation on the field hospital, the group of victims on the top of the plot are characterized by long presence times in the Examination room, and they correspond to the victims coded green, and the patients that spent more time in the Ambulance triage room are the more serious cases, coded yellow and red.

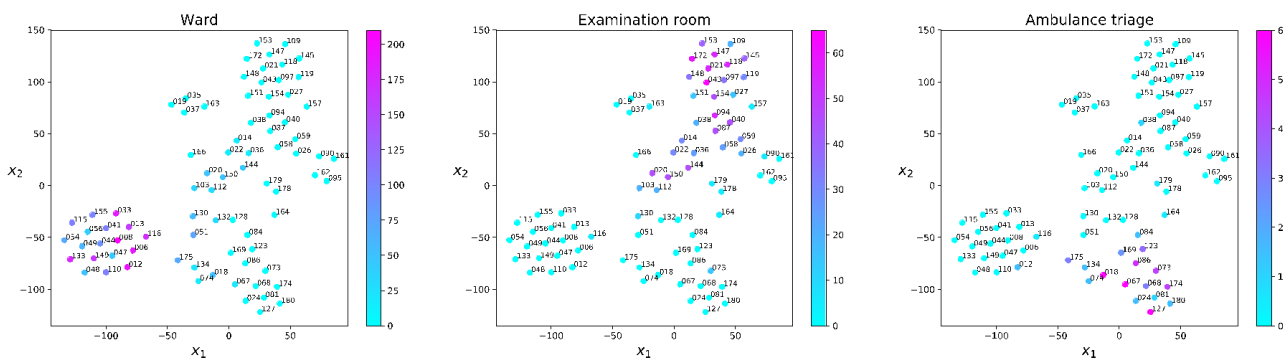

Figure 8. Presence patterns of individual victims in hospital rooms visualized using t-Distributed Stochastic Neighbor Embedding (t-SNE) to map high-dimensional patient vectors to a 2-dimensional feature space ($X_1$ and $X_2$ axes). Each point corresponds to a patient. Victims with similar presence vectors are mapped to neighboring points in the plane. Points (patients) are colour coded according to the time spent in minutes in "Ward" (left panel), "Examination room" (middle panel) and "Ambulance triage" room (right panel).

# Discussion

## Principal findings

With the present study, we report the first quantitative assessment of social contact patterns in live MCI simulation, based on wearable proximity sensors. Our study showed the feasibility of the use of wearable proximity sensors to measure contact patterns during a MCI functional exercise. We obtained simple charts and contact matrices which allow for direct visualization of potentially missed opportunities for improvement of the MCI response.

Major events, from natural, anthropic or terrorism, have become an increasingly frequent reality, Hospital disaster preparedness (HDP) with an effective medical response has therefore taken on increased importance. Simulation can provide a safe learning environment while replicating the chaotic environment typical of an actual disaster, with multiples victims and different gradient of severity of injuries [2]. Planning and realizing a mass casualty functional exercise requires significant time and resources [22, 23]. Therefore, an accurate and objective method for evaluation of MCI simulations is needed, as the functional exercises represent a link between disaster planning and disaster response [6]. Although after-exercise debriefing sessions, during which participants discuss deficiencies warranting improvement, are routinely conducted, there is no commonly used and validated method for evaluating health performance during MCI exercises.

In this study, we used proximity sensors to evaluate the contacts between individuals and the flow of the victims during an MCI simulation. The system provides information on the caregiver-casualty relations and flow of victims coherent with the severity of the diagnosis. We obtained contact matrices based on median cumulative time spent in proximity between victims with different triage score and different ISS and the rescuers. Our results showed that there were no differences between the median time passed by medical doctors and nurses with victims with different triage, both in Pre-hospital and Hospital area. At the scene of the accident, this result is consistent with the chaotic environment typical of a disaster, where the medical staff task allocation is challenging, and the medical interventions are equally distributed between patients with different severity injuries. However, significant differences between time in proximity by medical doctors and nurses with victims with different ISS were observed at Pre-hospital area. Both caregiver categories spent more time in contact with patients with the most severe injuries (classified as Critical), and in particular, with patients classified as Moderate in term of ISS. When presented and discussed critically at the debriefing it turned out that the majority of Moderate victims suffered bone fractures and due to this the immobilization and stabilization procedures required a lot of time before being mobilized and transferred to the field hospital. At Hospital area, nurses spent higher time in contact with Moderate and Serious patients as ISS.

Although there are no significant differences between the time passed by medical doctors and nurses with victims with different triage, proximity sensors revealed that medical doctors and nurses spent relevant time with patients with Red and Yellow triage, both in Pre-hospital and in Hospital area. Minor wounded (green codes) were predominantly managed by EMS staff on scene and HCA in hospital, allowing medical doctors and nurses to spend more time with the patients in most need. The quantitative measurement of contact patterns provided the opportunity to debate about it during the debriefing and identifying further strategies and counter actions. In disaster and MCI environments coordination for task allocation is challenging. The analysis of temporal features of contact links between caregivers and casualties revealed proportional resource utilization of different healthcare skills for different victim severity triages and ISS codes.

## Contact data and network analysis

We obtained aggregated contact networks of the participants involved in the simulation and we calculated the mean degree of the networks (*i.e.*, the mean number of connections between participants) for the whole duration of the simulation. Our results showed that the average degree was similar both for in-Hospital and Pre-hospital area. The mean number of connections is lower for medical doctors and nurses compared with those of EMS and HCA personnel. In other words, the medical staff interacted with a low number of patients by focusing treatment on a limited number of cases (see Multimedia Appendix). Moreover, the contact networks showed a high heterogeneity of the cumulative time spent in proximity (*e.i.*, weight of the edges) between participants, despite the short duration of the simulation. Our results show a highly heterogeneous distribution of contact durations characterized by a heavy tail, this outcome confirms the presence of 'universal' feature of contacts patterns with most contacts of short duration and few long-lasting contacts. A similar general distribution of contact durations has been observed in other settings, including schools [20], hospitals [10], and households [24]. Moreover, the density of the networks (*i.e.*, fraction of all possible edges that are present in the network) was calculated, in order to study the topology of networks built for each caregivers' categories and victims. The networks are sparse in both scenarios, in particular for EMS and HCA. However, we found that the density varied through the severity of injuries of the victims for the medical doctors and nurse, that showed a higher number of potential connections with victims with more serious conditions (see Multimedia Appendix). The analysis of temporal evolution of the number of contacts between participants revealed a high concentration of contacts during the middle part of the simulation at the Pre-hospital area, from 8 PM to 9:30 PM, even after the transfer of part of victims to the field hospital. The peak at the end of simulation is most likely due to an artefact: the meeting of participants shortly before the collections of sensors. On the Hospital area the number of contacts was stationary until the transfer of patients from Pre-hospital to Hospital area, then the number gradually increased as expected (see Multimedia Appendix).

## Flow and presence patterns of victims in the hospital

The deployment of sensors inside the hospital allowed to study the casualty flow. This analysis enabled to evaluate whether the patients were correctly headed by the healthcare personnel, consistently to the severity of the diagnosis and the expected location for such diagnosis, in other words if patients with high acuity pathology were correctly occupying high acuity areas of the hospital and vice-versa if low acuity patients were managed without wasting precious resources. Our results showed that the presence of patients in the hospital rooms were consistent with triage code and diagnosis. Green victims spent most of their time in examination rooms, Walk-in-triage room and Waiting room. Six Red victims out of ten spent time in the Resus, (*i.e.*, resuscitation area). Three Red patients spent most of their time in Intensive care (*i.e.*, Intensive Care Unit), and their diagnosis included head and chest trauma, and septic shock. Eleven Yellow patients with minor injuries and stable trauma out of fourteen spent time in Ambulance triage. We studied the potential presence of bottlenecks for the rooms where victims were examined by medical staff and they were waiting to receive treatments (Ambulance triage, Examination room, Waiting room, and Walk-in-triage), and evident presence of bottlenecks was not found. However, in our study, the number of victims in the field hospital is limited, and further evaluations of bottlenecks in simulations with greater number of victims are necessary.

Similar presence patterns of victims coded with the same triage were observed in the rooms of field hospital. In particular, the more serious cases (yellow and red codes) spent more time in the Ambulance triage room and the victims coded green spent more time in the Examination room. Our results showed a correspondence between the triage of the victims and the treatment given to the patients from the point of view of the permanence times in the rooms of the field hospital, and the interactions with the caregivers (see Multimedia appendix). In other words, victims with comparable severity injuries were managed in a similar way in the field hospital. The red victim 172 was an exception in this trend: the ideal triage code of the victim 172 was yellow. This result is consistent with the high degree of over-triage of yellow casualties as red occurred in this simulation. It is well known that over-triage may cause fatigue of staff, depletion of resources, and impairment of efficient flow of critically injured patients through the system to definitive care [25]. The use of physical space plays a key role in managing a sudden influx of injured people or patients [26]. The evaluation of casualty flow and hospital space usage during exercises is a necessary first step in disaster preparedness and readiness by hospital authorities.

## Limitations

It is important to highlight some limitations of the present study. The exercise was organized as realistically as possible. Despite this, it is still a simulation, and the patients could portray only a limited number of changes in clinical condition. Another potential issue concerns the possibility that participants changed their behaviour because they were wearing sensors and knew they were participating in a scientific measure.

However, the methods presented in this paper can be useful to detect contact patterns in the very specific context of MCI, thus allowing implementing tailored prevention strategies accordingly.

The measurement approach we used here has also limitations. Contacts were defined as face-to-face proximity, but no information on the possible occurrence of a physical contact between the two individuals is available [11], and consequently no information on the interactions caregiver – casualty is provided. Moreover, the short period of time of data collection (3 hours) also limits the ability to draw conclusion on what happens at longer time scales. However, through the use of the proximity sensing platform, long-time studies are allowed.

## Conclusions

In conclusion, our study shows that using wearable sensors based on proximity sensors technology is feasible for obtaining a precise measurement of the pattern of close contacts among individuals during an MCI simulation. It represents, to our knowledge, the first example of unsupervised data collection of face-to-face contacts during an MCI exercise by means of wearable proximity sensors. The unsupervised measurement of contact patterns with proximity sensors provides a unique opportunity to monitor the interactions between participants without the involvement of direct observers, which could impair the exercise realism. Moreover, the analysis of contact patterns may help to identify specific interactions between health staff – patient, in order to evaluate the decisions taken and the performance as the task allocation. In this study, the use of the sensors as fixed devices allowed to analyse the casualty flow in the field hospital, in order to assess the use of physical space and resources allocation. The versatility of the system makes it possible to repeat similar studies in different environments and to compare results across contexts. Future works could include comparison of contact patterns on different settings of mass casualty simulations, in order to improve medical process, resource utilization and decision making.

## Acknowledgments

This study was supported by the Lagrange Project of the ISI Foundation funded by the CRT Foundation to LO, LG, MQ, MT, AP, KK and CC.

LC, PLI, DC, FDC and CC designed the study. LO, LG, CC and MQ analysed the data. LO and LG drafted the manuscript. All authors critically revised and edited the draft and approved of the final version. LO, LG and CC had full access to all the data in the study and take responsibility for the integrity of the data and the accuracy of the data analysis.

**Multimedia appendix 1**

**Degree distributions of caregivers**

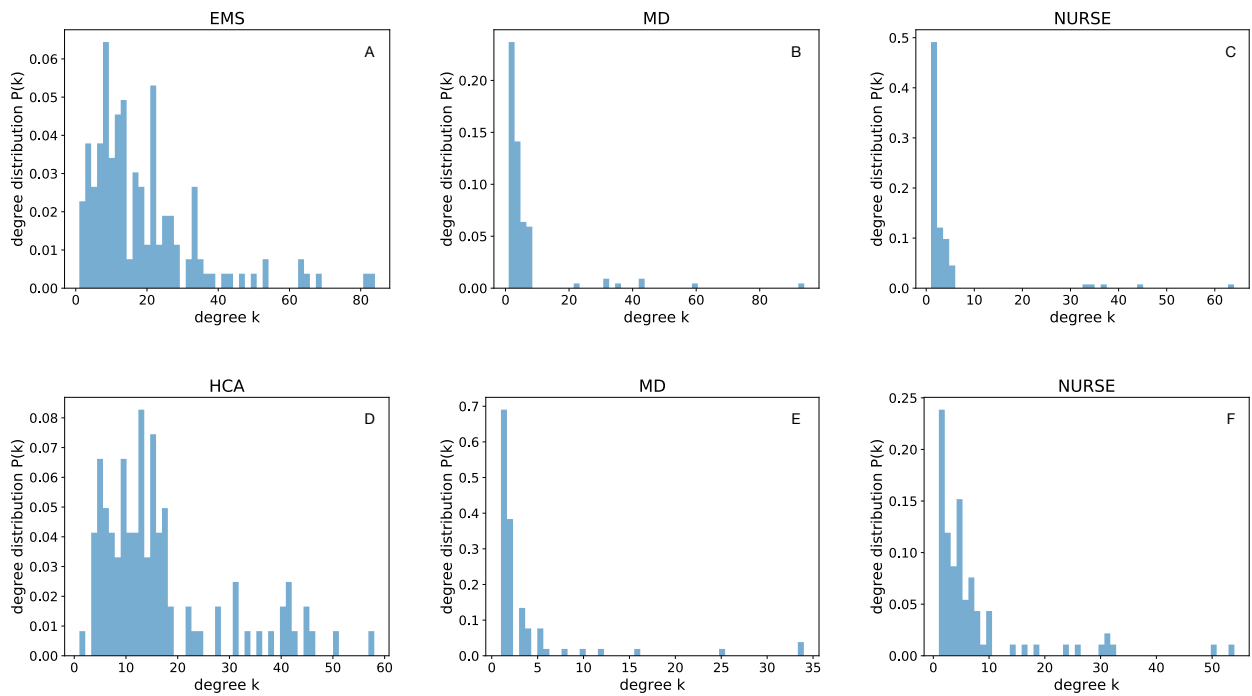

Figure S1. Degree distribution of caregivers: EMS (panel A) on Pre-hospital area; Medical doctors on Pre-hospital area (panel B) and Hospital area (panel E); Nurses on Pre-hospital area (panel C), and Hospital area (panel F); HCA on Hospital area (panel D)

The mean degree of EMS was ⟨k⟩ = 18.36, median = 14 (range 1- 84) on Pre-hospital area (panel A). The mean degree of medical doctors was ⟨k⟩ = 6.12, median = 3 (range 1 – 94) on scene of the accident (panel B), and ⟨k⟩ = 3.52, median = 2 (range 1 -34) on field hospital (panel E). The mean degree of the nurses was ⟨k⟩ = 4.08, median = 2 (range 1 – 64) (panel C), and ⟨k⟩ = 7.52, median = 5 (range 1 – 54) (panel F) on Pre-hospital and Hospital area respectively. The mean degree of HCA was ⟨k⟩ = 16.37, median = 13 (range 1- 58) on Hospital area (panel D).

**Multimedia appendix 2**

*Contacts temporal evolution*

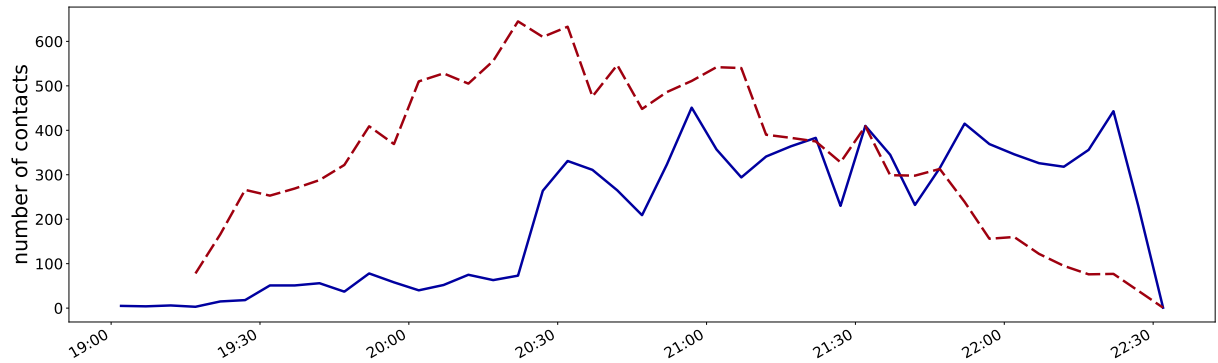

Figure S2. Temporal evolution of the number of contact on Pre-hospital area (red dashed line) and on Hospital area (blue line)

On Pre-hospital area the number of contacts gradually increased until 8:30 PM, and the largest number of contacts occurred between 8:30 PM and 9:05 PM; then the contacts gradually decreased. On Hospital area the number of contacts were stationary until 8:25, then the contacts gradually increased until 10:20 PM. At 8:15 PM patients started to be transferred from Pre-hospital to in-Hospital area.

**Multimedia appendix 3**

*Networks density*

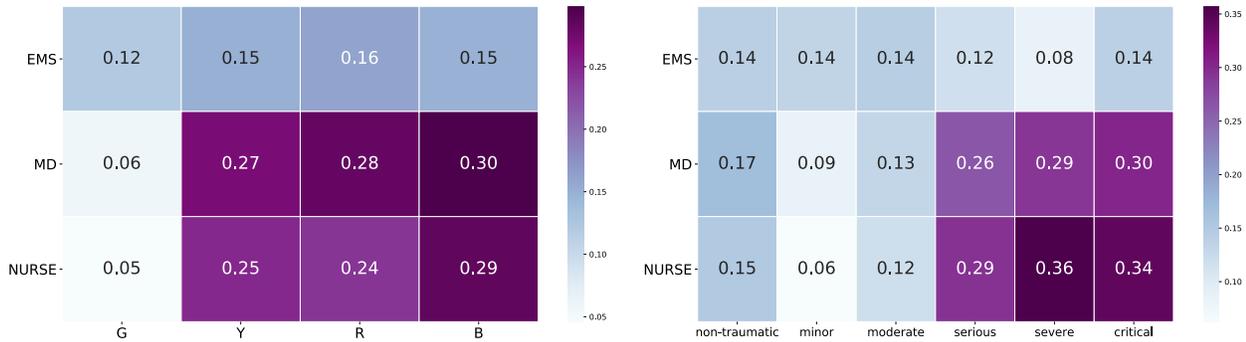

Figure S3. Density of networks in Pre-hospital area. Density of networks (*i.e.*, fraction of all possible edges that are present in the network) built for each caregivers' categories and victims with different triage (right panel), and different ISS (right panel).

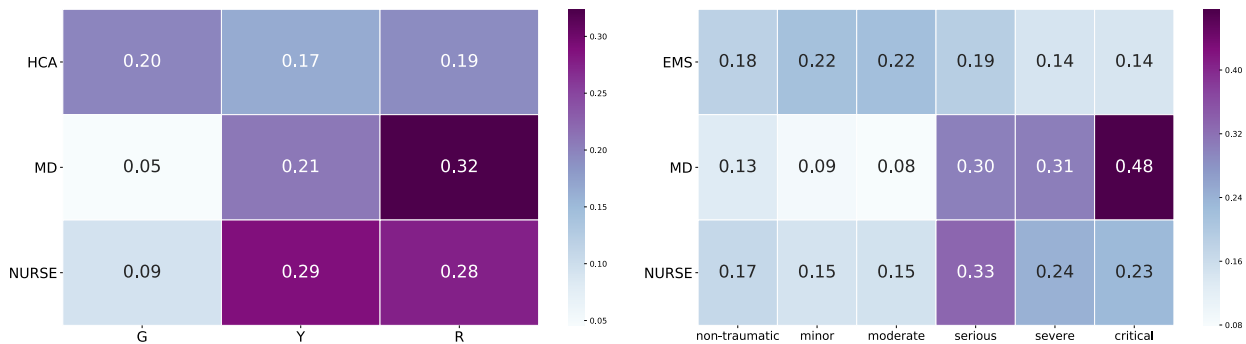

Figure S3. Density of networks in Hospital area. Density of networks (*i.e.*, fraction of all possible edges that are present in the network) built for each caregivers' categories and victims with different triage (right panel), and different ISS (right panel).

**Multimedia appendix 4**

*Cumulative time and percentage of time spent in contact*

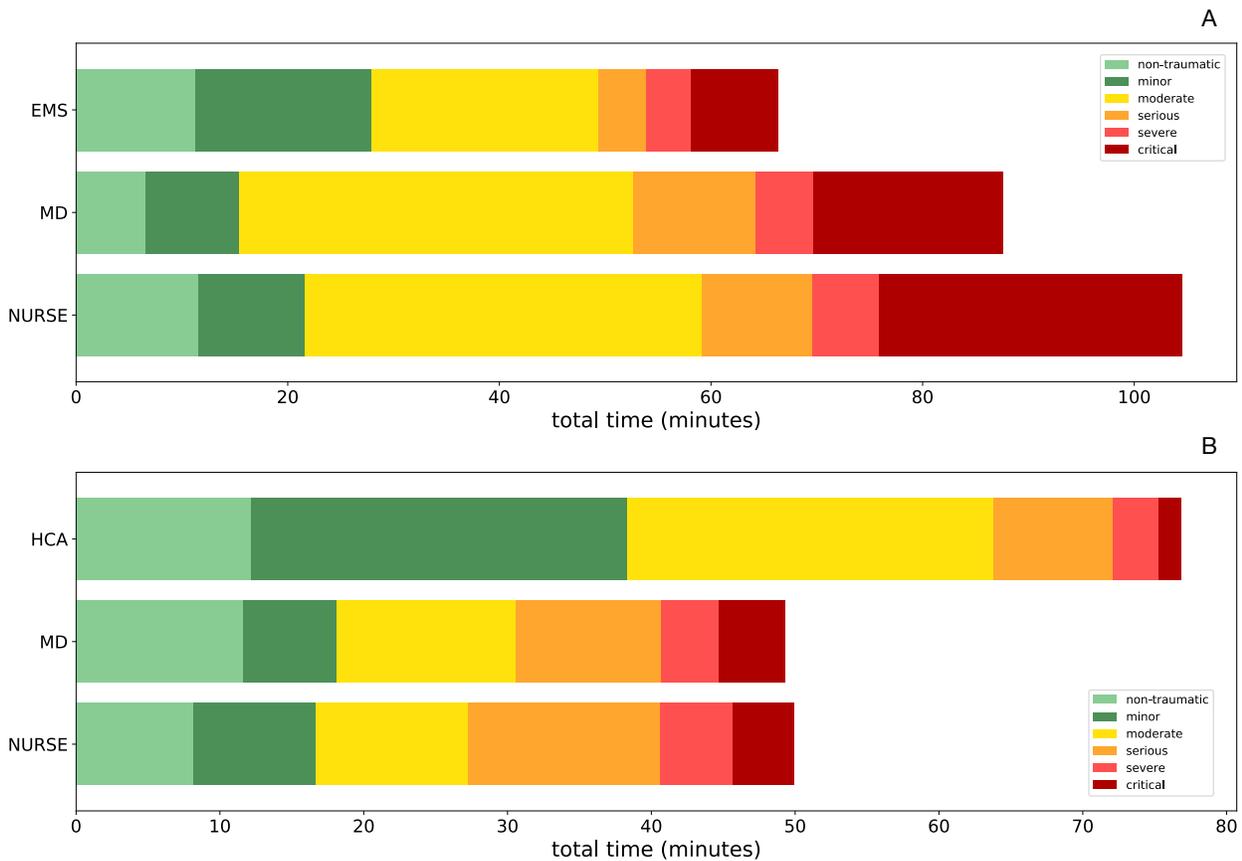

Figure S5. Cumulative time in contact (normalized on total number of participants belonging to each caregiver category) between caregivers and victims with different ISS at Pre-hospital area (panel A) and Hospital area (panel B).

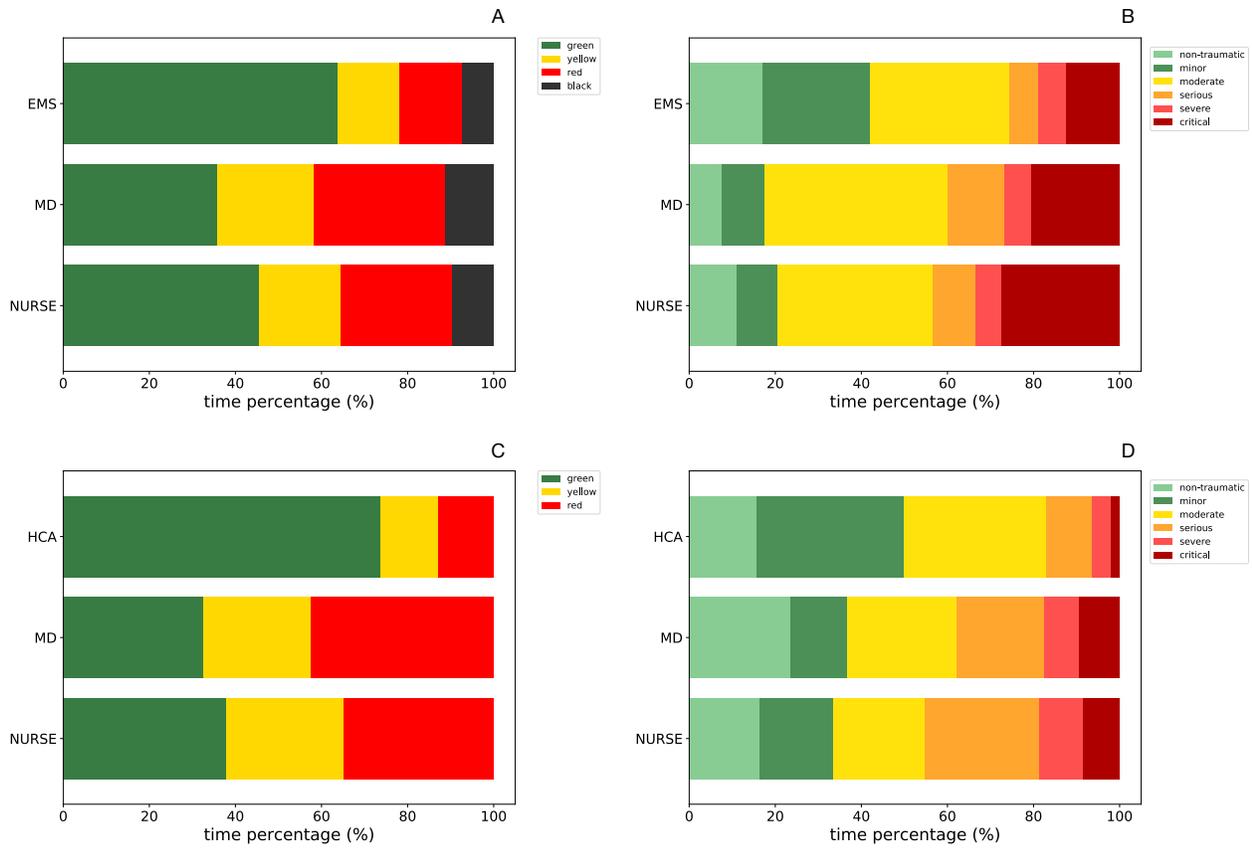

Figure S6. Percentage of time in contact between caregivers and victims with different triage and ISS at Pre-hospital area (panel A, and panel B) and Hospital area (panel C, and panel D).

**Multimedia appendix 5**

*Spatial and interaction patterns of victims in the Hospital*

We studied the spatial and interaction patterns of individual victims in hospital. For each patient we build a feature vector containing the time spent by that patient in each of the 15 hospital rooms, and the time spent in contact with caregivers (medical doctors, nurses and HCA). The resulting set of 18-dimensional vectors (one per patient) is visualized using a dimensionality reduction technique known as t-Distributed Stochastic Neighbor Embedding (t-SNE), that maps each 18-dimensional patient vector to a 2-dimensional feature space ($X_1$ and $X_2$ axes).

Figure S7. Spatial and interaction patterns of individual victims in hospital visualized using t-Distributed Stochastic Neighbor Embedding (t-SNE) to map high-dimensional patient vectors to a 2-dimensional feature space ($X_1$ and $X_2$ axes). Each point corresponds to a patient. Victims with similar presence vectors are mapped to neighboring points in the plane. Victims are colour coding by triage code (left panel), and by the start location of the simulation (red: accident site; blue: field hospital).